\documentclass[aps,prl,twocolumn,nofootinbib,superscript address]{revtex4-2}
\usepackage[utf8]{inputenc}
\usepackage[english]{babel}
\usepackage[T1]{fontenc}
\usepackage{amsfonts}
\usepackage{amsmath}
\usepackage{physics}
\usepackage{hyperref}
\usepackage{xcolor}
\usepackage{graphicx}

\newcommand{\wbar}{\overline{W}}
\newcommand{\hg}{\hbar\Gamma_\mathrm{tot}}
\newcommand{\kt}{k_\mathrm{B} T}

\addto\captionsenglish{}

\begin{document}
\title{Extra cost of erasure due to quantum lifetime broadening}

\author{Joe Dunlop} \email[he/him ]{j.dunlop@exeter.ac.uk} \affiliation{Physics and Astronomy, University of Exeter, Exeter EX4 4QL, United Kingdom}

\author{Federico Cerisola}
\affiliation{Physics and Astronomy, University of Exeter, Exeter EX4 4QL, United Kingdom}

\author{Juliette Monsel}
\affiliation{Department of Microtechnology and Nanoscience (MC2), Chalmers University of Technology, S-412 96 Göteborg, Sweden}

\author{Sofia Sevitz}
\affiliation{Institut für Physik, Potsdam University, 14476 Potsdam, Germany}

\author{Jorge Tabanera-Bravo}
\affiliation{Mathematical Biophysics Group, Max Planck Institute for Multidisciplinary Sciences, Göttingen 37077, Germany}

\author{Jonathan Dexter}
\affiliation{Dept. of Materials, University of Oxford, Oxford OX1 3PH, United Kingdom}

\author{Federico Fedele}
\affiliation{Department of Engineering Science, University of Oxford, Parks Road, Oxford, OX1 3PJ, United Kingdom}

\author{Natalia Ares}
\affiliation{Department of Engineering Science, University of Oxford, Parks Road, Oxford, OX1 3PJ, United Kingdom}

\author{Janet Anders} \affiliation{Physics and Astronomy, University of Exeter, Exeter EX4 4QL, United Kingdom} \affiliation{Institut für Physik, Potsdam University, 14476 Potsdam, Germany}
\date{\today}

\begin{abstract}
The energy cost of erasing a bit of information was fundamentally lower bounded by Landauer, in terms of the temperature of its environment: $W\geq k_\mathrm{B} T \ln 2$. However, in real electronic devices, the information-bearing system is usually in contact with two or more electrodes, with different temperatures and chemical potentials. It is not clear what sets the cost of erasure in such nonequilibrium situations. One promising technology for testing the thermodynamic limits of information processing is quantum dots, in which a bit is encoded in the presence or absence of a single electron.  We here develop a thermodynamic description of devices of this type and find that, in addition to the electrode temperatures, the potential difference across the quantum dot and lifetime broadening of its energy level contribute to the minimum work cost of erasure. In practical contexts, these contributions may significantly outweigh the cost due to temperature alone.
\end{abstract}
\maketitle

The well-known fundamental limit on the energy consumption of information processing devices is Landauer's bound, which holds that the process of erasing a bit of information must dissipate at least $\kt \ln 2$ of heat, where $T$ is the temperature of the thermal environment \cite{landauer_61}. As a keystone of the connection between thermodynamics and information theory, the foundational status of the bound has long been discussed \cite{bennett_03,maroney_09}. Now, basic limits on energy dissipation are a pressing practical problem: information technology consumes around five per cent of the global electricity supply \cite{gupta_21}, and thermal management is a primary bottleneck for integrated circuit performance \cite{Ogrenci_15}. Part of the solution must come from optimising the physical implementation of basic logic operations. As current CMOS transistor switching-energies are in the hundreds of $\kt$ \cite{datta_22}, improvement will require analysis and mitigation of the factors which prevent the Landauer bound from being approached in real-world conditions.

Progress towards this end has included proof-of-concept experimental demonstrations of erasure at $\kt$-scale energy costs (more recently in solid state electronics) \cite{exp_toyabe_10,exp_berut_12,exp_jun_14,exp_koksi_14,exp_roldan_14,exp_hong_16,exp_gaudenzi_18}. Meanwhile, theoretical advances have accounted for constraints beyond temperature \cite{basnett_13}, including the effects of finite-speed driving, finite-sized reservoirs, strong coupling, quantum coherence and limited control complexity \cite{reeb_14,goold_15,miller_20,vu_22,rolandi_23,taranto_24}, leading to proposed optimisations using techniques from thermodynamic geometry \cite{zhen_21,proesmans_20,proesmans_20_2,diana_13,scandi_19,abiuso_20}. However, important considerations have been overlooked: in particular, electronic components almost never interact with a single homogeneous thermal environment.

In this Letter, we analyse the thermodynamics of erasure in a quantum dot charge bit, which exchanges electrons with two electrodes with different temperatures and chemical potentials. Quantum dots represent the limit of miniaturisation for devices with a source, gate and drain electrode, and they are regarded as a promising platform for low-energy information processing \cite{patel_21,klupfel_16,tilke_01}. By explicitly considering optimal erasure protocols, we identify and bound the scale of three inherent sources of energy dissipation. In addition to recovering the Landauer bound, we find independent contributions to the energy cost resulting from lifetime broadening of the dot’s energy level, and source-drain potential difference. With reference to existing experimental devices \cite{vigneau_22,gustavsson_06}, we find that these contributions can outweigh temperature-related dissipation in realistic regimes of operation, sometimes to the extent that Landauer’s bound is practically irrelevant. Finally, we discuss the extent to which these energy costs might be mitigated, the generalisability of the results to other device types, and the theoretical significance of the role of lifetime broadening as a quantum source of noise.

\begin{figure}
    \centering
    \includegraphics[width=1\columnwidth]{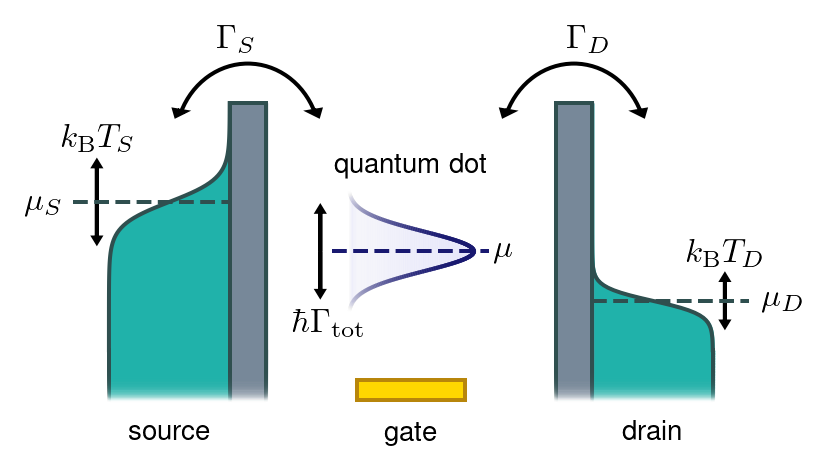}
    \caption{Schematic energy level diagram of a quantum dot charge bit device. The dot exchanges electrons with a source and drain electrode via quantum tunnelling, with characteristic rates $\Gamma_S$ and $\Gamma_D$ respectively. Electrons in the source and drain are described by the Fermi-Dirac distribution with chemical potentials $\mu_S,\mu_D$, and temperatures $T_S,T_D$ respectively. A bit of information is encoded in the electronic occupation $p$ of a single level of the quantum dot at energy $\mu$, which is subject to lifetime broadening with scale $\sim \hbar \Gamma_\mathrm{tot}$. The occupation of the dot may be manipulated by externally varying $\mu$ via the electrostatic field from a gate electrode.}
    \label{device}
\end{figure}

\textit{Model Device}---We consider the charge bit device depicted in Fig.~\ref{device}, consisting of a single-level quantum dot which exchanges electrons via quantum tunnelling with a source and drain electrode. The dot's average electronic occupation $0\,{<}\,p\,{<}\,1$ varies as a function of its energy level $\mu$, which in turn is controlled by the electrostatic field from a gate electrode. The presence or absence of an electron in the dot may be taken to represent a $0$ or $1$ digit, encoding a bit of information. In this context, erasure means a transformation from the state of maximum ignorance $(p=\frac{1}{2})$ to certainty about the occupation of the dot, which can mean either $p=0$ or $p=1$ (reset to the logical zero or one state respectively).


\begin{figure}
    \centering
    \includegraphics[width=0.92\columnwidth]{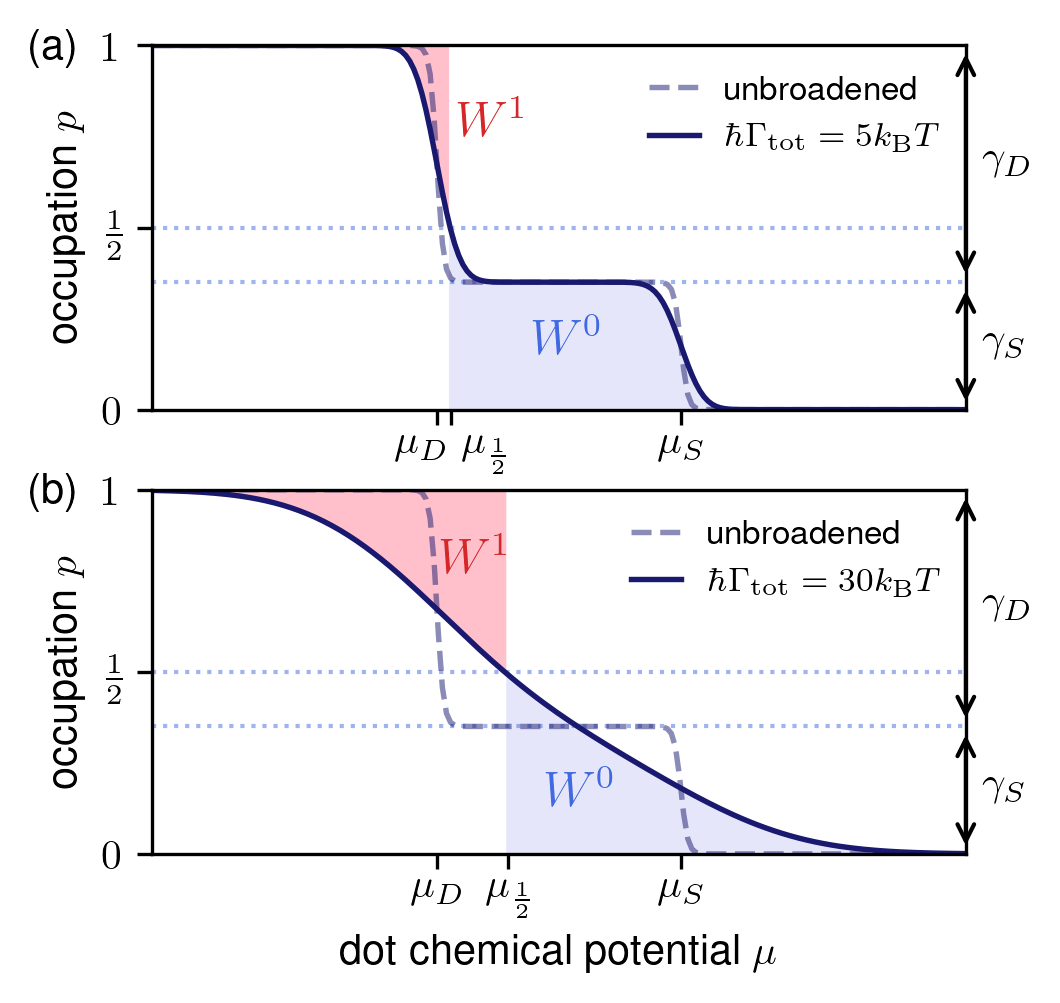}
    \caption{
    Average occupation $p$ plotted against energy level $\mu$ for a quantum dot in simultaneous contact with two electron reservoirs. The occupation is described by a weighted sum of the reservoirs' Fermi-Dirac distributions (dashed curve, both panels), which is smoothed due to lifetime broadening with scale $\hbar\Gamma_\mathrm{tot}$. Starting at $p{=}\frac{1}{2}$, the minimum work required to prepare the $p{=}0$ state is represented by the blue shaded area $W^0$; and the cost of preparing $p{=}1$ is represented by the red area $W^1$. Both processes represent erasure of a bit of information. In panel \textbf{(a)}, the source-drain chemical potential difference $(\mu_S-\mu_D)$ is considerably larger than either the thermal ($k_\mathrm{B} T$) or lifetime broadening ($\hbar\Gamma_\mathrm{tot}$) energy scales, and therefore the average work cost of erasure $\wbar=\frac{1}{2}(W^0{+}W^1)$ is approximately $\frac{1}{2}\gamma_S (\mu_S - \mu_D)$, i.e. bias-dominated. In panel \textbf{(b)}, lifetime broadening is comparable to the bias, and contributes significantly to $\wbar$. 
    In both panels, the temperatures $T_S {=} T_D {=} T$, the bias $\mu_S-\mu_D = 36 k_\mathrm{B} T$, and the tunnelling ratios $\gamma_S \,{=}\, 0.35,\,\gamma_D \,{=}\, 0.65$. Lifetime broadening is taken to have Gaussian form with standard deviation $\hbar\Gamma_\mathrm{tot}$.
    }
    \label{occupation_fig}
\end{figure}

The work cost of erasure depends on the occupation distribution $p(\mu)$, which we model using a rate equation, see Supplemental Material (SM) \cite{supp}. We assume that the dot exchanges electrons with the source and drain at fixed rates $\Gamma_{S,D}$; and that the electron reservoirs are described by the Fermi-Dirac distributions $f_{S,D}(\varepsilon)$ with respective temperatures $T_{S,D}$ and chemical potentials $\mu_S\geq \mu_D$. If lifetime broadening is neglected, the steady-state occupation of the dot is described by a convex combination of the source and drain Fermi-Dirac distributions, weighted by the tunnelling ratios $\gamma_{S,D} = \frac{\Gamma_{S,D}}{\Gamma_S + \Gamma_D}$ \cite{beenakker_92,supp}:
\begin{equation}\label{unbroadened}
    p_0(\mu) = \gamma_S\; f_S(\mu) + \gamma_D\; f_D(\mu).
\end{equation}
However, at higher tunnelling rates, the stronger coupling to the leads allows new transitions of electrons between the dot and the reservoirs. This results in an effective  broadening of the available electronic states in the leads, which is commonly taken 
to have Lorentzian or Gaussian form, centred on $\mu$ and with characteristic width $\hbar\Gamma_\mathrm{tot}$, where $\Gamma_\mathrm{tot} = \Gamma_S + \Gamma_D$ \cite{meerwaldt_12,haldar_24}.
%
%
The overall occupation $p$ (plotted in Fig.~\ref{occupation_fig}) is then given by the \textit{cross-correlation} of the broadening distribution $g$ with the unbroadened occupation $p_0$ from Eq.~\eqref{unbroadened}:
\begin{equation}\label{broadened}
    p(\mu) = (g \star p_0)(\mu) = \int_{-\infty} ^ \infty g(\varepsilon-\mu)p_0(\varepsilon)d\varepsilon.
\end{equation}


\textit{Optimal Erasure Protocols}---The quantum dot's occupation can be manipulated by externally varying its energy level $\mu$ via the gate electrode. The rate of work done on the dot is defined by $\dot{W} = p\, \dot{\mu}$ \cite{alicki_79}. Let $\mu_\frac{1}{2}$ denote the value of $\mu$ such that $p(\mu_\frac{1}{2})=\frac{1}{2}$. We here outline a thermodynamically reversible protocol for erasure to logical zero. Starting at $p=\frac{1}{2}$, the energy level $\mu$ is quasistatically\footnote{i.e. over a timescale much longer than $\Gamma_\mathrm{tot}^{-1}$.} raised from $\mu_\frac{1}{2}$ towards $+\infty$ such that the dot's occupation vanishes. The level is then reset to $\mu_\frac{1}{2}$ much faster than the dot's equilibration timescale, such that the final occupation remains at $p=0$. The overall work done on the dot in this process is $W^0 = \int_{\mu_\frac{1}{2}}^{\infty} p(\mu)d\mu$.
Erasure to `one' proceeds similarly. The energy level is slowly lowered from $\mu_\frac{1}{2}$ towards $-\infty$, before being quickly raised back to $\mu_\frac{1}{2}$, preserving the occupation $p=1$. The work cost in this case is $W^1 = \int_{-\infty}^{\mu_\frac{1}{2}} \left(1-p(\mu)\right)d\mu$. For more details, see SM~\cite{supp}.

Aside from exceptionally symmetric cases\footnote{such as at zero bias ($\mu_S = \mu_D$), or if the source and drain electrode have equal temperatures and tunneling rates.}, $W^0$ and $W^1$ differ from one another (plotted as shaded areas in Fig.~\ref{occupation_fig}). In the context of information processing, ones are required just as frequently as zeros for efficient coding \cite{shannon}. A fair measure of the cost of state reset is the average, $\wbar = \frac{1}{2}(W_0 + W_1)$, which may be written as \cite{supp}:
\begin{equation}\label{wbar}
    \wbar = \frac{1}{2}\int_{-\infty}^\infty \big|\mu - \mu_\frac{1}{2}\big| \left(-\frac{dp}{d\mu}\right)d\mu.
\end{equation}
$\wbar$ is a tight lower bound on the average work cost, since the above protocols are reversible.
Moreover, $\wbar$ is associated solely with information erasure rather than any net change in the dot's internal energy $\mu p$: the change in $\mu p$ is eliminated by averaging, since both protocols reset the energy level to $\mu_\frac{1}{2}$.

If $p(\mu)$ is interpreted as a complementary cumulative distribution function, then Eq.~\eqref{wbar} is equivalent to half the mean absolute deviation about the median $\mu_\frac{1}{2}$. This provides the basis to use properties of the mean absolute deviation to disentangle contributions to $\wbar$ from different physical parameters, as well as a heuristic that the work cost of erasure relates directly to the spread of the occupation distribution.


\textit{Bounds on the Work Cost of Erasure}---While Eq.~\eqref{wbar} is straightforward to integrate numerically, it is not possible to obtain a closed formula for $\wbar$ due to the difficulty of inverting $p(\mu)$ for $\mu_\frac{1}{2}$. Short of an exact formula, it will be informative to place analytic bounds on $\wbar$. We here present an illustrative overview of limiting cases to motivate such a bound; a formal derivation is available in SM \cite{supp}.

For a quantum dot in contact with a \textit{single} electrode with negligible lifetime broadening, $p(\mu)$ is equal to the Fermi-Dirac distribution, and its spread is characterised solely by the reservoir temperature $T$: in this case, the familiar Landauer bound is recovered from Eq.~\eqref{wbar}, with $\wbar = \kt \ln 2$. A similar result extends to the two-reservoir case, provided that source-drain potential bias is also negligible, such that $p(\mu)$ is described by Eq.~\eqref{unbroadened} with $\mu_\frac{1}{2} {=} \mu_S {=} \mu_D$. By linearity, Eq.~\eqref{wbar} reduces to $\wbar = k_\mathrm{B} (\gamma_S T_S + \gamma_D T_D) \ln 2$, a version of the Landauer bound involving the \textit{average} of the reservoir temperatures, weighted by the tunnelling ratios $\gamma_{S,D}$. Let us denote this thermal energy scale as $E_\mathrm{therm}$:
\begin{equation}\label{etherm}
    E_\mathrm{therm} = k_\mathrm{B} (\gamma_S T_S + \gamma_D T_D) \ln 2.
\end{equation}

If, instead, source-drain potential bias is the dominant energy scale, such that $\kt$ and $\hg$ are negligible in comparison to $\mu_S {-} \mu_D$, then $p(\mu)$ is effectively a sum of step functions at $\mu_S$ and $\mu_D$, with a plateau at $p = \gamma_S$ in between. This is approximately the situation in Fig.~\ref{occupation_fig}a. Supposing that $\gamma_S < \gamma_D$ (i.e. faster particle exchange with the drain than source), then $\mu_\frac{1}{2} \approx \mu_D$, and erasure to the $p=1$ state is comparatively cheap (of the order $\kt$ or $\hg$). On the other hand, erasure to $p=0$ involves raising the energy level from $\mu_D$ to $\mu_S$ at near-constant occupation $p=\gamma_S$, so that $W^0 \approx \gamma_S (\mu_S - \mu_D)$. By a mirroring argument, if instead $\gamma_S > \gamma_D$, then $W^1 \approx \gamma_D (\mu_S - \mu_D)$ and $W^0$ is negligible. Generally, then, in bias-dominated regimes, the average cost of erasure is approximated by a characteristic \textit{bias energy scale}, $\wbar \approx E_\mathrm{bias}$, given by:
\begin{equation}\label{ebias}
    E_\mathrm{bias} = \tfrac{1}{2}\min\{\gamma_S,\gamma_D\}(\mu_S - \mu_D).
\end{equation}

Thirdly, we can consider the limit where lifetime broadening dominates. Using a general property of the cross-correlation, the mean absolute deviation $D$ of the occupation distribution $p = g\star p_0$ can be bounded in terms of that of the unbroadened distribution $p_0$ and broadening function $g$, as follows: $\max\{D(g),D(p_0)\}\leq D(p) \leq D(g) + D(p_0)$ (see SM \cite{supp} for the derivation). If the broadening is such that $D(g) \gg D(p_0)$, which is the case if $\hg \gg \max\{E_\mathrm{therm},E_\mathrm{bias}\}$, then the average work cost of erasure is effectively set by the mean absolute deviation of the broadening function, $\wbar\approx \frac{1}{2} D(g)$. We will label this the \textit{lifetime broadening energy scale}:
\begin{equation}\label{ebroad}
    E_\mathrm{broad} = \tfrac{1}{2} D(g) \,\equiv\, \tfrac{1}{2}\int_{-\infty}^\infty \left| \varepsilon - m_g \right| g(\varepsilon)d\varepsilon,
\end{equation}
where $m_g$ is the median of $g$. The exact dependence of $E_\mathrm{broad}$ on the tunnelling rates depends on the form of the broadening distribution: for example if $g$ is a Gaussian with standard deviation $\hg$, then $E_\mathrm{broad} = \frac{\hg}{\sqrt{2\pi}}$. If $g$ is Lorentzian with scale $\hbar\Gamma_\mathrm{tot}$, then the mean absolute deviation diverges, implying an unbounded energy cost for perfect erasure. However, approximate erasure to a occupation within $\eta$ of $0$ or $1$ is still possible, with work cost $E_\mathrm{broad}^\eta = \frac{\hbar\Gamma_\mathrm{tot}}{2\pi} \ln[\sec^2(\pi(\tfrac{1}{2}-\eta))]$
, as shown in SM \cite{supp}.



\begin{figure}
    \centering
    \includegraphics[width=1\columnwidth]{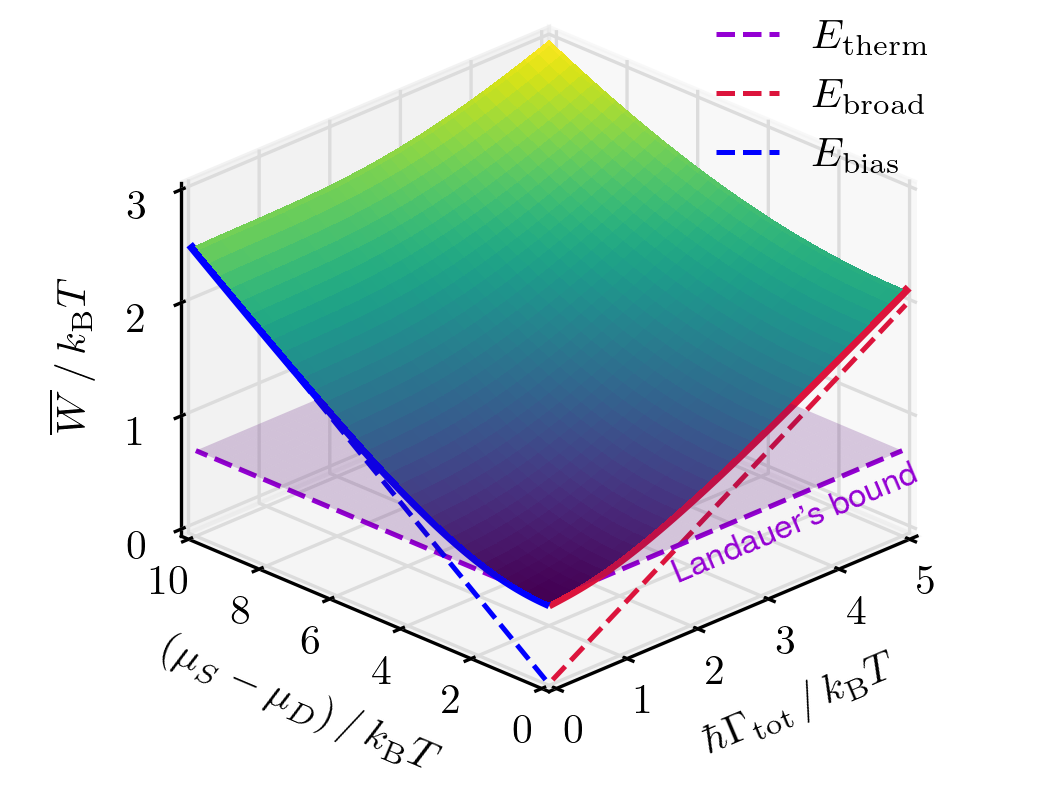}
    \caption{
    The optimal average work cost of erasure $\wbar$ for a quantum dot in contact with two electrodes, plotted against source-drain bias ($\mu_S - \mu_D$) and lifetime broadening ($\hbar\Gamma_\mathrm{tot}$), at fixed temperature $T_S {=} T_D {=} T$. At zero bias and zero broadening, an ideal erasure process can saturate the Landauer bound, $\wbar = k_\mathrm{B}T \ln 2$. However, if either $\mu_S - \mu_D$ or $\hbar\Gamma_\mathrm{tot}$ exceeds a few times $k_\mathrm{B} T$, then the work cost is significantly larger. In general, $\wbar$ is bounded from below by the largest out of the three characteristic energy scales $E_\mathrm{therm}$, $E_\mathrm{broad}$ and $E_\mathrm{bias}$ as defined in Eqs.(\ref{etherm},\ref{ebias},\ref{ebroad}), which are plotted here as dotted lines. 
    In this plot, the source and drain tunneling rates are taken to be equal: $\Gamma_S{=}\Gamma_D{=}\frac{\Gamma_\mathrm{tot}}{2}$.
    }
    \label{3d_plot}
\end{figure}

We have identified three independent energy scales, $E_\mathrm{therm}, E_\mathrm{bias}$ and $E_\mathrm{broad}$, each of which emerges as the minimum average work cost of erasure $\wbar$ in the case that the other two vanish. This is the first main result of the paper. It remains to treat the more general scenario where temperature, bias and lifetime broadening all contribute nontrivially. As shown in SM \cite{supp}, $\wbar$ can be bounded as follows:
\begin{equation}\label{bound}
    \max\left\{ E_\mathrm{therm}, E_\mathrm{bias}, E_\mathrm{broad} \right\} \leq \wbar \leq E_\mathrm{therm} + E_\mathrm{bias} + E_\mathrm{broad}.
\end{equation}
This is the second main result. By the left-hand inequality, the average work cost of erasure cannot under any circumstances be made smaller than any of the three contributing energy scales as defined in Eqs. (\ref{etherm},\ref{ebias},\ref{ebroad}). The meaning of the right-hand inequality is more subtle. Certainly there is no upper limit to how much energy might be dissipated in an erasure operation when performed inefficiently. However, $\wbar$ relates to an ideal, thermodynamically reversible state reset. The bound here implies that, in the absence of further constraints, there is no principle preventing erasure at an average cost equal to or less than $E_\mathrm{therm} + E_\mathrm{bias} + E_\mathrm{broad}$. 

In general, neither inequality is tight. However, the upper bound on $\wbar$ is never more than threefold greater than the lower, ensuring a correct order-of-magnitude estimate in all parameter regimes. The window becomes much narrower if any one of the three characteristic energy scales dominates. Figure \eqref{3d_plot} compares $\wbar$ against the bounds imposed by $E_\mathrm{therm}, E_\mathrm{bias}$ and $E_\mathrm{broad}$.

In experimental contexts, direct measurement of energy dissipation at microscopic scales is extremely difficult. 
The present approach provides a way to assess thermodynamic performance using more readily accessible measurements (temperature, voltage and tunnelling rates). Given these parameters, Eq.~\eqref{bound} reduces the estimation of $\wbar$ to a back-of-the-envelope calculation; and a more precise value is possible by integrating Eq.~\eqref{wbar}. Moreover, by separately quantifying the scale of contributions to $\wbar$, the approach may be used identify thermodynamic bottlenecks, a crucial step towards mitigating the cost.


\textit{Discussion}---We have analysed the thermodynamics of erasure in a model of a quantum dot charge bit, incorporating a near-ubiquitous feature of current information processing devices -- that the information-bearing system is in contact with two electrodes with a potential difference. This marks a qualitative difference from the standard approach: the system is inherently out of thermodynamic equilibrium, and can at best occupy a dynamical steady state.

Landauer's bound can give the impression that the work cost of erasure can be arbitrarily low if carried out in a cold enough environment. Our results show that other factors can dominate dissipation at low temperatures. Since quantum dot devices often operate in the sub-kelvin regime, it is of practical importance to obtain a tighter bound.
We find that $k_B T$ no longer represents a fixed information-energy exchange rate, with the conversion instead determined by a non-linear combination of temperatures, chemical potentials and lifetime broadening.
The additional costs are intrinsic to the source-gate-drain architecture, and unavoidable even in the limit of perfect quasistatic operation. The charge bit considered here encodes information in a single electron, and a larger penalty can reasonably be expected where information is stored redundantly in multiple microscopic degrees of freedom.

Far from a marginal correction, non-Landauer terms are the dominant component of erasure cost in some real experimental devices. For example, Ref.~\cite{vigneau_22} details a device with operating parameters $T_{S,D} = 40\,\mathrm{mK}$, $\mu_S {-} \mu_D = 200\,\mathrm{\mu eV}$, and $\Gamma_S = 6.3\,\mathrm{GHz},\,\Gamma_D = 250\,\mathrm{GHz}$. Here, the characteristic energy scales are $E_\mathrm{therm} = 2.4\,\mathrm{\mu eV},\, E_\mathrm{bias} = 2.5\,\mu\mathrm{eV}$ and $\,E_\mathrm{broad} = 67\,\mathrm{\mu eV}$; and integrating \eqref{wbar} gives $\wbar = 68\,\mathrm{\mu eV}$ (assuming Gaussian broadening). $\wbar$ is primarily determined by lifetime broadening in this case, with temperature and bias affecting only the second significant figure.
By contrast, the device in Ref.~\cite{gustavsson_06} is bias-dominated: here $T_{S,D} = 350\,\mathrm{mK}$, $\mu_S {-} \mu_D = 500\,\mathrm{\mu eV}$, $\Gamma_S = 1.75\,\mathrm{kHz}$, and $\Gamma_D = 1.45\,\mathrm{kHz}$, 
for which $E_\mathrm{therm} = 21\,\mathrm{\mu eV},\, E_\mathrm{bias} = 113\,\mathrm{\mu eV}$ and $\,E_\mathrm{broad} = 8.4\times 10^{-7}\,\mathrm{\mu eV}$.
Lifetime broadening is negligible here due to the slow tunnelling rates, and thermal broadening contributes to $\wbar$ in the third significant figure, with $\wbar = 117\,\mathrm{\mu eV}$.

The present paper only accounts for the work done by the field controlling the quantum dot's energy level, neglecting the energy dissipated when electrons flow from source electrode to drain (which in some cases may be partially recovered as a source of useful work \cite{wen_20,tabanera_24}). The cost of maintaining potential difference, in addition to the distinct bias-related contribution in Eq.~\eqref{ebias}, might be mitigated by reducing voltage across the device. 
Likewise, the broadening-related cost (Eq. \ref{ebroad}) could be suppressed by reducing the tunnelling rates between the quantum dot and electrodes. However, this leads to a compromise in the maximum possible speed of erasure, since the timescale for the dot to respond to changes in gate voltage is $\sim \frac{1}{\Gamma_\mathrm{tot}}$. This principle is related to Bremermann's limit \cite{bremermann_67,deffner_20}, and distinct from the dissipation which occurs when driving is fast in comparison to the equilibration rate \cite{zhen_21}. This represents a practical thermodynamic consequence of 
the effective energy-time uncertainty relation \cite{wigner_88,busch_08}. A similar penalty due to lifetime broadening would arise for a quantum dot which exchanges electrons with a single reservoir electrode\footnote{This can be seen by taking one of the tunnelling rates to 0.}, such as in a single-electron box \cite{pekola_13}. 


While the present work has considered information encoded in the charge of the quantum dot, a device operating on the same principles may be used as a single-electron transistor, switching source-drain current via Coulomb blockade.
A key question for future work is whether lifetime broadening and potential difference have a similar influence on energy dissipation in the switching process; as well as how the effects scale to logic gates.

\textit{Acknowledgements}---
We would like to thank M. T. Mitchison, K. Aggarwal and J. Parrondo for stimulating discussions.
J. Dunlop acknowledges funding from the Engineering and Physical Sciences Research Council (EP/T518049/1).
FC and NA acknowledge funding from the Foundational Questions Institute Fund (FQXi-IAF19-01).
JM acknowledges funding from the Knut and Alice Wallenberg foundation.
SS and JA acknowledge support from Deutsche Forschungsgemeinschaft (DFG 384846402).
JT-B acknowledges support from the Alexander von Humboldt Foundation.
J. Dexter acknowledges support from the EPSRC.
FF and NA acknowledge funding from the European Union Quantum Flagship project ASPECTS (Grant Agreement No. 101080167).
NA acknowledges funding from the European Research Council (ERC) under the European Union's Horizon 2020 research and innovation program (Grant Agreement No. 948932).
JA acknowledges funding from EPSRC (EP/R045577/1). JA thanks the Royal Society for support.
Views and opinions expressed are those of the authors only and do not necessarily reflect those of funding bodies.



%

\end{document}


\title{Supplemental Material: Extra cost of erasure due to quantum lifetime broadening}

\author{Joe Dunlop} \email[he/him ]{j.dunlop@exeter.ac.uk} \affiliation{Physics and Astronomy, University of Exeter, Exeter EX4 4QL, United Kingdom}

\author{Federico Cerisola}
\affiliation{Physics and Astronomy, University of Exeter, Exeter EX4 4QL, United Kingdom}

\author{Juliette Monsel}
\affiliation{Department of Microtechnology and Nanoscience (MC2), Chalmers University of Technology, S-412 96 Göteborg, Sweden}

\author{Sofia Sevitz}
\affiliation{Institut für Physik, Potsdam University, 14476 Potsdam, Germany}

\author{Jorge Tabanera-Bravo}
\affiliation{Mathematical Biophysics Group, Max Planck Institute for Multidisciplinary Sciences, Göttingen 37077, Germany}

\author{Jonathan Dexter}
\affiliation{Dept. of Materials, University of Oxford, Oxford OX1 3PH, United Kingdom}

\author{Federico Fedele}
\affiliation{Department of Engineering Science, University of Oxford, Parks Road, Oxford, OX1 3PJ, United Kingdom}

\author{Natalia Ares}
\affiliation{Department of Engineering Science, University of Oxford, Parks Road, Oxford, OX1 3PJ, United Kingdom}

\author{Janet Anders} \affiliation{Physics and Astronomy, University of Exeter, Exeter EX4 4QL, United Kingdom} \affiliation{Institut für Physik, Potsdam University, 14476 Potsdam, Germany}
\date{\today}

\maketitle

\setlength{\parskip}{1em}
\setlength{\parindent}{0em}

\section{Electron tunneling rate equation and steady-state occupation}




We first briefly note that, while the paper exclusively discusses a single-level quantum dot, all results should hold to a very good approximation for an arbitrary $N-$level dot, provided that the spacing of energy levels is much larger than the source-drain potential difference or broadening energy scales. In that case, at most one energy level (say, the $j^\mathrm{th}$) lies within the bias window and is partially occupied, with the all lower levels fully occupied and all higher levels completely empty at all times. In that case, the bit is encoded in the occupation of the $j^{th}$ level, and all others can be neglected.

We assume that the electronic occupation of the source and drain electrodes are described by the Fermi-Dirac distributions $f_{S/D} (\varepsilon) = \frac{1}{1+\exp(\beta_{S/D} (\varepsilon - \mu_{S/D}))}$. The tunnelling rates between the quantum dot and electrodes are denoted $\Gamma_S$ and $\Gamma_D$, and assumed to be constant. Then, at a given energy $\varepsilon$, the electron currents $\Gamma$ in and out of the dot depend on the dot occupation $p$ and the electrode occupations $f_{S}$ and $f_{D}$ are as follows:
\begin{equation}\label{rates}
    \begin{split}
        \Gamma^\mathrm{in}_S &= \Gamma_S f_S(\varepsilon)\, (1-p(\varepsilon)),\\\Gamma^\mathrm{out}_S &= \Gamma_S (1-f_S(\varepsilon))\, p(\varepsilon),\\
        \Gamma^\mathrm{in}_D &= \Gamma_D f_D(\varepsilon)\, (1-p(\varepsilon)),\\
        \Gamma^\mathrm{out}_D &= \Gamma_D (1-f_D(\varepsilon))\, p(\varepsilon).
    \end{split}
\end{equation}
The steady-state occupation $p_0$ is achieved when the total currents in and out are balanced:
\begin{equation}\label{integraldecomp}
    \begin{split}
        \Gamma^\mathrm{in}_S + \Gamma^\mathrm{in}_D &= \Gamma^\mathrm{out}_S + \Gamma^\mathrm{out}_D\\
        &\implies p_0(\varepsilon) = \tfrac{\Gamma_S}{\Gamma_S + \Gamma_D} f_S(\varepsilon) + \tfrac{\Gamma_D}{\Gamma_S + \Gamma_D} f_D(\varepsilon).
    \end{split}
\end{equation}
Introducing the shorthand $\gamma_{S/D} = \tfrac{\Gamma_{S/D}}{\Gamma_S + \Gamma_D}$, we arrive at Eq. (1) in the main text, noting that $\gamma_S + \gamma_D = 1$. In fast-tunnelling regimes, the energy level of the quantum dot is described by a distribution $g$ about the chemical potential $\mu$, due to lifetime broadening. Then the overall (broadened) occupation of the dot is given by averaging the unbroadened occupation $p_0(\varepsilon)$, weighted by the probability $g(\varepsilon\, |\, \mu)$ that the dot's energy level takes the value $\varepsilon$ given that the chemical potential is $\mu$:
\begin{equation}\label{given}
    p(\mu) = \int_{-\infty} ^ \infty g(\varepsilon\, |\,\mu)p_0(\varepsilon)d\varepsilon.
\end{equation}
If we additionally assume that the shape of the broadening distribution $g$ is independent of the dot's chemical potential, such that we can write $g(\varepsilon\,|\,\mu) = g(\varepsilon -\mu)$, then the above leads to Eq. (2) (this assumption is revisited in Eqs \eqref{dg} onwards). In this case, the broadened occupation $p(\mu)$ is equivalent to the \textit{cross-correlation} $(g \star p_0)(\mu)$, a product of functions closely related to convolution:
\begin{equation}\label{xcor}
    p(\mu) = g\star p_0 (\mu) = \int_{-\infty} ^ \infty g(\varepsilon\, -\,\mu)p_0(\varepsilon)d\varepsilon.
\end{equation}

\section{Accounting for work}

Suppose that the dot's chemical potential is driven  as $\mu(t)$, and that its occupation varies as $p(t)$. We take the rate at which work is done on the dot by the driving field to be given by \cite{alicki_79}:
\begin{equation}\label{workrate}
    \tfrac{dW}{dt} = p(t) \tfrac{d\mu}{dt}.
\end{equation}
If the driving is much slower than the dot occupation's equilibration rate (i.e. the tunnelling rates), then the occupation is given by the instantaneous steady state $p(\mu(t))$ as given in Eq. (2). In this quasistatic regime, the explicit time-dependence can be eliminated from Eq. \eqref{workrate}, which integrates to:
\begin{equation}\label{quaswork}
    W = \int_{\mu_i}^{\mu_f} p(\mu) d\mu,
\end{equation}
where $\mu_i$ and $\mu_f$ are the initial and final values of the dot's chemical potential. In the other extreme, when the chemical potential is suddenly \textit{quenched} from $\mu_i$ to $\mu_f$ over a timescale much quicker than the tunnelling rates, then the occupation undergoes negligible evolution during the process. If the occupation is initially in the steady state $p(\mu_i)$, then the work done on the dot during the quench is:
\begin{equation}\label{quenchwork}
    W = \int_{\mu_i}^{\mu_f} p(\mu_i) d\mu = (\mu_f-\mu_i)p(\mu_i).
\end{equation}

\textbf{Erasure to Zero---}Starting in the configuration $(\mu=\mu_{\frac{1}{2}},p=\frac{1}{2})$, the dot level is quasistatically raised to some $\mu_\mathrm{hi} \gg \mu_S$, for which the steady state occupation is vanishingly small: without lifetime broadening, $p(\mu_\mathrm{hi})\sim e^{-\beta(\mu_\mathrm{hi} - \mu_S)}$. The new configuration is $(\mu=\mu_\mathrm{hi},p=p(\mu_\mathrm{hi}))$, and by Eq. \eqref{quaswork}, the work cost of raising the energy level is $\int_{\mu_\frac{1}{2}}^{\mu_\mathrm{hi}} p(\mu) d\mu$. Then, the level is quenched back to $\mu_\frac{1}{2}$, so that the final configuration is $(\mu=\mu_\frac{1}{2},p=p(\mu_\mathrm{hi}))$. The (negative) work done on the dot during the quench is $-(\mu_\mathrm{hi}-\mu_\frac{1}{2})p(\mu_\mathrm{hi})$, by Eq.\eqref{quenchwork}. Overall, in the limit as $\mu_\mathrm{hi}\to +\infty$, the operation takes $(\mu=\mu_{\frac{1}{2}},p=\frac{1}{2}) \mapsto (\mu=\mu_{\frac{1}{2}},p=0)$, at a total work cost given by:
\begin{equation}
    \begin{split}
        W^0 &= \lim_{\mu_\mathrm{hi}\to \infty} \left[\int_{\mu_\frac{1}{2}}^{\mu_\mathrm{hi}} p(\mu) d\mu - (\mu_\mathrm{hi}-\mu_\frac{1}{2})p(\mu_\mathrm{hi})\right]
        = \int_{\mu_\frac{1}{2}}^{\infty}p(\mu) d\mu,
    \end{split}
\end{equation}
assuming that $p(\mu)$ decays sufficiently fast.

\textbf{Erasure to One---}Again, starting at $(\mu=\mu_{\frac{1}{2}},p=\frac{1}{2})$, the dot level is slowly lowered to $\mu_\mathrm{lo} \ll \mu_D$ so that the new configuration is $(\mu=\mu_\mathrm{lo},p=p(\mu_\mathrm{lo}))$ with $p(\mu_\mathrm{lo}) \approx 1$; during which the (negative) work done on the dot is $W = -\int_{\mu_\mathrm{lo}}^{\mu_\frac{1}{2}} p(\mu) d\mu$. Then, the energy level is quickly quenched back to $\mu_\frac{1}{2}$ at a work cost $(\mu_\frac{1}{2} - \mu_\mathrm{lo})p(\mu_\mathrm{lo})$. Taking the limit as $\mu_\mathrm{lo}\to-\infty$, the protocol maps $(\mu=\mu_{\frac{1}{2}},p=\frac{1}{2}) \mapsto (\mu=\mu_{\frac{1}{2}},p=1)$, at a work cost given by:
\begin{equation}
    \begin{split}
        W^1 &= \lim_{\mu_\mathrm{lo}\to -\infty} \left[(\mu_\frac{1}{2}-\mu_\mathrm{lo}) p(\mu_\mathrm{lo}) - \int^{\mu_\frac{1}{2}}_{\mu_\mathrm{lo}} p(\mu) d\mu\right]
        = \int^{\mu_\frac{1}{2}}_{-\infty}\left(1-p(\mu)\right) d\mu.
    \end{split}
\end{equation}

\textbf{Thermodynamic reversibility---}Both the erasure-to-zero and erasure-to-one protocols are reversible in the following sense. If the initial configuration was $(\mu=\mu_{\frac{1}{2}},p=0)$ and $\mu$ was quenched to $\mu_\mathrm{hi}$ and then gradually lowered back to $\mu_{\frac{1}{2}}$ (the time-reverse of the erasure-to-zero protocol), then the configuration would be mapped to $(\mu=\mu_{\frac{1}{2}},p=\frac{1}{2})$ at a work cost $-W^0$. Therefore, if the erasure-to-zero protocol was applied to $(\mu=\mu_{\frac{1}{2}},p=\frac{1}{2})$ and \textit{immediately} followed by its time-reverse, then the configuration would be unchanged, and the total work cost would vanish. The same is true for erasure to one. The caveat is that neither $(\mu=\mu_{\frac{1}{2}},p=0)$ nor $(\mu=\mu_{\frac{1}{2}},p=1)$ are steady-state configurations, and if $\mu$ were held fixed following the erasure protocols, the dot population would begin to irreversibly equilibrate back towards $\frac{1}{2}$.

Resetting the energy level to $\mu_\frac{1}{2}$ following erasure ensures that there is no average change in the dot's internal energy $U = \mu p$. In the erasure-to-zero operation $(\mu = \mu_\frac{1}{2}, p=\frac{1}{2})\mapsto (\mu = \mu_\frac{1}{2}, p=0)$, the change in internal energy is $\Delta U = -\frac{1}{2}\mu_\frac{1}{2}$, and in erasure-to-one $(\mu = \mu_\frac{1}{2}, p=\frac{1}{2})\mapsto (\mu = \mu_\frac{1}{2}, p=1)$, the change is $\Delta U = \frac{1}{2}\mu_\frac{1}{2}$. Averaging with equal weighting gives $\overline{\Delta U}=0$. Moreover, if the quantum dot has additional energy levels outside the bias window, resetting $\mu$ means that there is no change in the internal energy associated with those levels.

\textbf{Relation to Mean Absolute Deviation---}Defining the average work cost $\wbar \equiv \frac{1}{2} \left(W^0+W^1\right)$, we have:
\begin{equation}\label{W_is_D}
    \begin{split}
        2\overline{W} &= \int_{\mu_{\frac{1}{2}}}^{\infty} p(\mu) d\mu + \int_{-\infty}^{\mu_{\frac{1}{2}}} \left(1 - p(\mu)\right) d\mu\\
        &= \left[(\mu-\mu_{\frac{1}{2}})p(\mu)\right]_{\mu_{\frac{1}{2}}}^\infty - \int_{\mu_{\frac{1}{2}}}^\infty (\mu-\mu_{\frac{1}{2}}) \frac{dp}{d\mu}\, d\mu
        + \left[(\mu-\mu_{\frac{1}{2}})(1 -p(\mu))\right]_{-\infty}^{\mu_{\frac{1}{2}}} - \int_{-\infty}^{\mu_{\frac{1}{2}}} (\mu-\mu_{\frac{1}{2}}) \left(-\frac{dp}{d\mu}\right) d\mu\\
        &= \int_{-\infty}^\infty \big|\mu - \mu_{\frac{1}{2}}\big| \left(-\frac{dp}{d\mu}\right)d\mu.
    \end{split}
\end{equation}
If $p_0$ is as in Eq. \eqref{integraldecomp} and $g$ is a well-defined probability density function, then $p = g\star p_0$ meets the definition of a complementary cumulative distribution function (i.e. monotone decreasing, $p\to 0$ as $\mu\to +\infty$, $p\to 1$ as $\mu\to -\infty$). The corresponding probability density function (PDF) is $p'(\mu) = -\frac{dp}{d\mu}$, with median equal to $\mu_\frac{1}{2}$. Equation \eqref{W_is_D} is equivalent to the mean absolute deviation (MAD) of that distribution: in general for a PDF $f(x)$ with median $m$, the MAD is given by:
\begin{equation}\label{medianmin}
\begin{split}
    D(f) &= \int_{-\infty}^\infty |x-m| f(x) dx
    \,\equiv\, \min_{y\in \mathbb{R}} \int_{-\infty}^\infty |x-y| f(x) dx.
\end{split}
\end{equation}
The property that the MAD is minimised about the median has a physical consequence: the mean cost of erasure\footnote{Assuming equal weighting of $W^0$ and $W^1$.} from an arbitrary initial steady state $(p,\mu)$ can never be lower than from $(\mu=\mu_{\frac{1}{2}},p=\frac{1}{2})$, even though $W^0$ and $W^1$ might radically differ.

\section{Bounding the mean absolute deviation}

We here state and prove two general properties of the MAD, which will later be applied to bound $\wbar$.

\textbf{Lemma 1.} Let $f$ and $g$ be probability density functions. Then the mean absolute deviation $D$ of the cross-correlation function $f\star g$ is bounded by:
\begin{equation}\label{L1}
    \begin{split}
    \mathrm{(i)} \hspace{3em} & D(f\star g) \geq \max\{D(f),D(g)\}\\
    \mathrm{(ii)} \hspace{3em} & D(f\star g) \leq D(f) + D(g).
    \end{split}
\end{equation}
\textbf{Proof.} Recall $f\star g = \int_{-\infty}^\infty f(y)g(y-x) dy = \int_{-\infty}^\infty f(y+x)g(y)dy$. Letting $m_{fg}$ be the median of $f\star g$, we have:
\begin{equation}
    \begin{split}
        D(f\star g) &= \int_{-\infty}^\infty |x-m_{fg}| \left(\int f(y+x)g(y) dy\right)dx\\
        &= \int \left(\int |x-m_{fg}| f(y+x) dx\right)g(y)dy\\
        &\geq \int D(f) g(y)dy\\
        &= D(f).
    \end{split}
\end{equation}
Since $f\star g (x) = g \star f (-x)$, and since the mean absolute deviation of a function $h(x)$ is equal to that of $h(-x)$, then by the above reasoning we also have $D(f\star g)\geq D(g)$, which completes the proof of 1(i). On the other hand to show the upper bound,
\begin{equation}
    \begin{split}
        D(f\star g) &= \int_{-\infty}^\infty |u-m_{fg}| f\star g(u) du\\
        & \leq \int |u-(m_f - m_g)| f\star g(u) du\\
        &= \int |u-(m_f - m_g)| \left(\int f(u+y)g(y)dy\right)du\\
        &= \int \left(\int |u-(m_f - m_g)| f(u+y) du\right) g(y) dy\\
        &= \int \left(\int |x-y-(m_f - m_g)| f(x) dx\right) g(y) dy\\
        &\leq \iint \left(|x-m_f| + |y-m_g|\right) f(x)g(y) dx dy\\
        &= \int \left(\int |x-m_f| f(x) dx\right)g(y)dy + \int \left(\int |y-m_g| g(y) dy\right)f(x)dx\\
        &= \int D(f) g(y)dy + \int D(g) f(x)dx\\
        &= D(f) + D(g),
    \end{split}
\end{equation}
where $m_f$ and $m_g$ are the medians of $f$ and $g$ respectively. In the first line, we used Eq.\eqref{medianmin}, and in the fifth line we used the substitution $x=u+y$. This completes the proof of 1(ii).

\textbf{Lemma 2.} Let $f$ be a probability density function which is even about its median $m_f$, such that for all $x$, $f(m_f+x)=f(m_f-x)$. Similarly, let $g$ be a probability density function which is even about its median $m_g$, where $m_g\leq m_f$. Let $p_f$ and $p_g$ be probabilities such that $p_f + p_g = 1$. 
Then the mean absolute deviation of the convex sum $p_f f + p_g g$ is bounded by:
\begin{equation}\label{L2}
    \begin{split}
        \mathrm{(i)} \hspace{3em} & D(p_f f + p_g g) \geq \max\Big\{ p_f D(\mu_f) + p_g D(\mu_g),\hspace{0.5em} \min\{p_f,p_g\} (m_f - m_g) \Big\}\\
        \mathrm{(ii)} \hspace{3em} & D(p_f f + p_g g) \leq p_f D(\mu_f) + p_g D(\mu_g)  +  \min\{p_f,p_g\} (m_f - m_g).
    \end{split}
\end{equation}

\textbf{Proof.} Let us denote $h(x) = p_f f(x) + p_g g(x)$; note that $h$ is also a probability density function and its median $m_h\in [m_g,m_f]$. We will use the fact that the median minimises the mean absolute deviation, in that for all $a$,  $\int_{-\infty}^\infty |x-a| f(x) dx \geq \int_{-\infty}^\infty |x-m_f| f(x) dx$. Then the mean absolute deviation of $h$ is bounded as follows: 
\begin{equation}\label{L2ia}
    \begin{split}
        D(h) &= \int_{-\infty}^\infty |x-m_h| (p_f f(x) +p_g g(x))dx\\
        &= p_f\int |x-m_h| f(x) dx + p_g\int |x-m_h|g(x) dx\\
        &\geq p_f\int |x-m_f| f(x) dx + p_g\int |x-m_g|g(x) dx\\
        &= p_f D(f) + p_g D(g).
    \end{split}
\end{equation}
On the other hand, we also have:
\begin{equation}\label{2bf}
    \begin{split}
        \int_{-\infty}^{\infty} |x-m_h|f(x)dx &= \int_{m_h}^{\infty} |x-m_h|  f(x) dx - \int_{-\infty}^{m_h} |x-m_h|  f(x) dx + 2 \int_{-\infty}^{m_h} |x-m_h| f(x) dx\\
        &= \int_{-\infty}^{m_f} (x-m_h) f(x) dx + \int_{m_f}^{\infty} (x-m_h) f(x) dx + 2 \int_{-\infty}^{m_h} |x-m_h| f(x) dx\\
        &= \int_{0}^{\infty} (m_f - u -m_h) f(m_f - u) du + \int_{0}^{\infty} (m_f + u -m_h) f(m_f + u) du\\
        &\hspace{15em} + 2 \int_{-\infty}^{m_h} |x-m_h| f(x) dx\\
        &= 2(m_f-m_h)\int_0^\infty f(m_f+u) du + 2 \int_{-\infty}^{m_h} |x-m_h| f(x) dx\\
        &= m_f - m_h + 2 \int_{-\infty}^{m_h} |x-m_h| f(x) dx.
    \end{split}
\end{equation}
In the penultimate line we used that $f(m_f + x) = f(m_f - x)$. In the final line we used that $\int_{m_f}^\infty f(x)dx = \frac{1}{2}$, by definition of the median. By a similar argument for $g$,
\begin{equation}\label{2bg}
 \int_{-\infty}^{\infty} |x-m_h|g(x)dx = m_h - m_g + 2 \int_{m_h}^{\infty} |x-m_h| g(x) dx.
\end{equation}
Combining \eqref{2bf} and \eqref{2bg}:
\begin{equation}\label{2bh}
\begin{split}
    D(h) = p_f (m_f - m_h) + p_g (m_h - m_g) + 2 \left( p_f \int_{-\infty}^{m_h} |x-m_h| f(x) dx + p_g \int_{m_h}^{\infty} |x-m_h| g(x) dx \right)
\end{split}
\end{equation}
Since the integral terms are always positive, it follows that:
\begin{equation}\label{L2ib}
\begin{split}
    D(h) &\geq p_f (m_f - m_h) + p_g (m_h - m_g)\\
    &\geq \min\{p_f,p_g\} (m_f - m_g).
\end{split}
\end{equation}
Taking inequalities \eqref{L2ia} and \eqref{L2ib} together, the claim 2(i) follows. Moreover, since $D(h) = \min_m \int_{-\infty}^\infty \abs{x-m} h(x) dx$, then from \eqref{2bh} we have: 
\begin{equation}\label{fonly}
    \begin{split}
        D(h) &= \min_m \left\{ p_f (m_f - m) + p_g (m - m_g) + 2 \left( p_f \int_{-\infty}^{m} |x-m| f(x) dx + p_g \int_{m}^{\infty} |x-m| g(x) dx \right) \right\}\\
        &\leq p_f (m_f - m_g) + p_g (m_g - m_g) + 2 p_f \int_{-\infty}^{m_g} |x-m_g| f(x) dx + 2 p_g \int_{m_g}^{\infty} |x-m_g| g(x) dx\\
        &= p_f (m_f - m_g) + p_g D(g) + 2 p_f \int_{-\infty}^{m_g} |x-m_g| f(x) dx\\
        &\leq p_f (m_f - m_g) + p_g D(g) + 2 p_f \int_{-\infty}^{m_f} |x-m_f| f(x) dx\\
        &= p_f (m_f - m_g) + p_g D(g) + p_f D(f).
    \end{split}
\end{equation}
In the penultimate line we used that $m_f \geq m_g$, and in the final line we used the assumption that $f$ is even about $m_f$. If we instead take $m=m_f$, then by the same arguments,
\begin{equation}\label{gonly}
    D(h) \leq p_g (m_f - m_g) + p_g D(g) + p_f D(f).
\end{equation}
Combining \eqref{fonly} and \eqref{gonly}, we certainly have $D(h) \leq \min\{p_f,p_g\} (m_f - m_g) + p_f D(f) + p_g D(g)$, which was the claimed bound 2(ii).

\section{Bounding the work cost of erasure}

We now apply Lemmas 1 and 2 to bound the work cost of erasure in the quantum dot device. Recall from Eq. \eqref{W_is_D} that $\wbar = \frac{1}{2} D(p')$, where $p'(\mu) = -\frac{dp}{d\mu}$. Combining equations \eqref{integraldecomp} and \eqref{xcor}, it can be seen that:
\begin{equation}\label{2w}
    \overline{W} = \frac{1}{2} D(g\star (\gamma_S f'_S + \gamma_D f'_D)),
\end{equation}
where $f'_\nu(\varepsilon) = -\frac{d}{d\mu} f_\nu (\varepsilon)$
, for $\nu = S,D$; which can be explicitly written as:
\begin{equation}\label{fprime}
    f'_\nu(\varepsilon) = \frac{\beta_\nu e^{\beta_\nu(\varepsilon - \mu_\nu)}}{(1 + e^{\beta_\nu(\varepsilon - \mu_\nu)})^2}.
\end{equation}
$f'_\nu$ defines a probability density function with median equal to the chemical potential $\mu_\nu$. 
The mean absolute deviation is given by:
\begin{equation}\label{Dfprime}
    D(f'_\nu) = \frac{2\ln2}{\beta_\nu}.
\end{equation}
Returning to \eqref{2w}, we can apply lemmas 1(i) followed by 2(i), using the fact that that $f'_\nu(\mu_\nu + \varepsilon) = f'_\nu(\mu_\nu - \varepsilon)$ for all $\varepsilon$:
\begin{equation}\label{belowbound}
    \begin{split}
        \overline{W} &\geq \frac{1}{2}\max\big\{D(g), \hspace{0.5em} D(\gamma_S f'_S + \gamma_D f'_D)\big\}\\
        &\geq \frac{1}{2}\max\big\{D(g), \hspace{0.5em} \gamma_S D(f'_S) + \gamma_D D(f'_D), \hspace{0.5em} \min\{\gamma_S,\gamma_D\}(\mu_S -\mu_D)\big\}\\
        &= \max\left\{\frac{1}{2} D(g), \hspace{0.5em} \left(\tfrac{\gamma_S}{\beta_S} + \tfrac{\gamma_D}{\beta_D} \right)\ln 2, \hspace{0.5em} \frac{1}{2}\min\{\gamma_S,\gamma_D\}(\mu_S -\mu_D)\right\}.
    \end{split}
\end{equation}
In the above, we have identified the three independent energy scales which contribute to the work cost of erasure:
\begin{equation}\label{energy_scales}
    \begin{split}
        E_\mathrm{therm} &= \left(\tfrac{\gamma_S}{\beta_S} + \tfrac{\gamma_D}{\beta_D} \right)\ln 2\\
        E_\mathrm{bias} &= \frac{1}{2}\min\{\gamma_S,\gamma_D\}(\mu_S -\mu_D)\\
        E_\mathrm{broad} &= \frac{1}{2} D(g).
    \end{split}
\end{equation}
Eq. \eqref{belowbound} is not a tight bound, and so far we cannot rule out that in fact an optimal erasure protocol would require much \textit{more} work that the right hand side of \eqref{belowbound} suggests. However, applying lemmas 1(ii) and 2(ii) to \eqref{2w}, we can upper-bound the minimum work cost:
\begin{equation}\label{abovebound}
    \begin{split}
        \overline{W} &\leq \frac{1}{2}\Big(D(g)\hspace{0.5em} + \hspace{0.5em}D(\gamma_S f'_S + \gamma_D f'_D)\Big)\\
        &\leq \frac{1}{2}\Big(D(g)\hspace{0.5em} + \hspace{0.5em} \gamma_S D(f'_S) + \gamma_D D(f'_D) \hspace{0.5em} + \hspace{0.5em} \min\{\gamma_S,\gamma_D\}(\mu_S -\mu_D)\Big)\\
        &= \frac{1}{2} D(g) \hspace{0.5em} + \hspace{0.5em} \left(\tfrac{\gamma_S}{\beta_S} + \tfrac{\gamma_D}{\beta_D} \right)\ln 2 \hspace{0.5em} + \hspace{0.5em} \frac{1}{2}\min\{\gamma_S,\gamma_D\}(\mu_S -\mu_D).
    \end{split}
\end{equation}
Putting this together with \eqref{belowbound}, we have the full bound:
\begin{equation}\label{Ebound}
    \max\Big\{ E_\mathrm{therm}, E_\mathrm{bias}, E_\mathrm{broad} \Big\} \leq \overline{W} \leq E_\mathrm{therm} + E_\mathrm{bias} + E_\mathrm{broad}.
\end{equation}

\section{Dependence on Lifetime Broadening Distribution}

We have established that the work cost of erasure is related to the mean absolute deviation of the lifetime broadening distribution, which in turn depends on the total tunnelling rate between quantum dot and electrodes. However, the exact relationship between $D(g)$ and $\Gamma_\mathrm{tot}$ is determined by the form of the broadening distribution, which in general depends on the details of the coupling between dot and electrodes. Here we will treat two common models: Lorentzian and Gaussian broadening.

\textbf{Gaussian broadening---}If lifetime broadening is described by a Gaussian $g(\varepsilon-\mu) = \frac{1}{\hg\sqrt{2\pi}}\exp(-\frac{1}{2}(\frac{\varepsilon-\mu}{\hg})^2)$, then the mean absolute deviation straightforwardly integrates as:
\begin{equation}
    D(g) = \frac{1}{\hg\sqrt{2\pi}}\int_{-\infty}^\infty |\varepsilon - \mu | \exp(-\frac{1}{2}\left(\frac{\varepsilon-\mu}{\hg}\right)^2) d\varepsilon = \hg\sqrt{\tfrac{2}{\pi}}.
\end{equation}
The lifetime broadening energy scale is then $E_\mathrm{broad} = \frac{1}{2}D(g) = \frac{\hg}{\sqrt{2\pi}}$.

\textbf{Lorentzian broadening---}If lifetime broadening is described by a Lorentzian distribution, then the work cost of perfect erasure to either $p=0$ or $p=1$ is unbounded, since the mean absolute deviation of a Lorentzian does not converge. However, we can derive an expression for work required to `erase' to $p$ within some small $\eta$-neighbourhood of $0$ or $1$.

Suppose that temperature and bias are negligible, and take $\mu_S = \mu_D = 0$ without loss of generality; such that the unbroadened distribution 
is described by a Heaviside step function $p_0(\mu) = H(-\mu)$. Take lifetime broadening to be described by a Lorentzian:
\begin{equation}\label{lorentzian}
    g(\epsilon - \mu) = \frac{\hbar \Gamma_\mathrm{tot}}{\pi \left[(\hbar\Gamma_\mathrm{tot})^2 + (\epsilon - \mu)^2\right]}.
\end{equation}
Then, using the symmetry of $g(\epsilon-\mu)$ about $\mu$, we have:
\begin{equation}
\begin{split}
    p(\mu) = g\star p_0 (\mu) &= \int_{-\infty}^\infty g(\varepsilon-\mu) H(-\varepsilon) d\varepsilon\\
    &= \int_{-\infty}^0 g(\varepsilon-\mu) d\varepsilon\\
    &= \int_{\mu}^\infty g(\varepsilon) d\varepsilon
\end{split}
\end{equation}
Let $0<\eta<1$, and let $\mu_\eta$ be such that $p(\mu_\eta) = \eta$. In fact $\mu_\eta$ is described by the quantile function for the Lorentzian:
\begin{equation}
    \mu_\eta = \hbar\Gamma_\mathrm{tot}\tan(\pi(\tfrac{1}{2}-\eta)).
\end{equation}
The work cost of raising $\mu$ from $\mu_\frac{1}{2} (= 0)$ to $\mu_\eta$ while the dot occupation remains in the steady state, followed by resetting the energy level from $\mu_\eta$ to $0$ at constant population $p=\eta$, is:
\begin{equation}
    \begin{split}
        W^\eta &= \int_0^{\mu_\eta} p(\mu) d\mu + \int_{\mu_\eta}^0 p(\mu_\eta) d\mu\\
        &= \left[\mu p(\mu)\right]_0^{\mu_\eta} - \int_0^{\mu_\eta} \mu \frac{dp}{d\mu} d\mu - \eta \mu_\eta\\
        &= \int_0^{\mu_\eta} \mu g(\mu) d\mu\\
        &= \frac{\hbar\Gamma_\mathrm{tot}}{2\pi}\left[\ln((\hbar\Gamma_\mathrm{tot})^2 + \mu^2)\right]_0^{\mu_\eta},
    \end{split}
\end{equation}
where we have used that $\frac{dp}{d\mu} = -g(\mu)$ in the third line. Substituting our expression for $\mu_\eta$:
\begin{equation}\label{wepsilon}
    \begin{split}
    W^\eta &= \frac{\hbar\Gamma_\mathrm{tot}}{2\pi} \left[ \ln((\hbar\Gamma_\mathrm{tot})^2 - \left(\hbar\Gamma_\mathrm{tot}\tan(\pi(\tfrac{1}{2}-\eta))\right)^2) -  \ln((\hbar\Gamma_\mathrm{tot})^2) \right]\\
    &= \frac{\hbar\Gamma_\mathrm{tot}}{2\pi} \ln[1+\tan^2(\pi(\tfrac{1}{2}-\eta))]\\
    &= \frac{\hbar\Gamma_\mathrm{tot}}{2\pi} \ln[\sec^2(\pi(\tfrac{1}{2}-\eta))].
    \end{split}
\end{equation}
So, while the energy cost of exact erasure diverges, for any given $\eta > 0$ the broadening-related work cost of erasing to within $\eta$ of $0$ or $1$ is proportional to $\hg$: the tunneling rates still determine the characteristic energy scale.

\textbf{Energy-dependent broadening---}
We assumed that the quantum dot's lifetime broadening distribution does not change shape depending on its chemical potential $\mu$ but instead is only shifted, allowing us to write $g(\varepsilon|\mu) = g(\varepsilon - \mu)$. we found that $p(\mu) = g\star p_0 (\mu)$ is a well-defined complementary cumulative distribution function (i.e. monotone-decreasing with $\lim_{\mu\to \infty}p(\mu) = 0$ and $\lim_{\mu\to -\infty}p(\mu) = 1$); and moreover that the mean absolute deviation $D(p)\geq D(g)$.

If we relax the assumption that $g(\varepsilon|\mu) = g(\varepsilon - \mu)$, and instead impose only that $\mu$ is the median of $g(\varepsilon|\mu)$, then $p(\mu)$ is given by Eq. \eqref{given},
and $D(g)$ is no longer a fixed quantity but depends on $\mu$:
\begin{equation}\label{dg}
    D(g)|_\mu = \int_{-\infty}^{\infty} |\varepsilon-\mu| \, g(\varepsilon|\mu) \, d\varepsilon.
\end{equation}
We might still hope to prove something like $D(p) \geq \min_y D(g)|_y$. However, this is not true at all. Let's consider the version of Eq. \eqref{dg} involving probability density functions $f(x),g(x|y)$:
\begin{equation}\label{not_xcor}
    h(y) = \int_{-\infty}^\infty f(x)\, g(x|y)\, dx.
\end{equation}
Here, $h(y)$ is not necessarily a well-defined probability density function, and $D(h)$ may vanish even if $f$ and $g$ are well-defined and have $D(f)>0$ and $D(g)|_y>0,\,\,\forall y$. 
In particular, $g(x|y)$ might be fine-tuned such that for all $y$, $g(x|y)$ and $f(x)$ have non-overlapping support.
A pathological example is as follows:
\begin{equation}
    \begin{split}
        f(x) &= \begin{cases} 1\,\,\mathrm{for}\,\,-\tfrac{1}{2}<x<\tfrac{1}{2}\\
        0\,\, \mathrm{elsewhere} \end{cases}\\
        g(x|y) &= \tfrac{1-\eta}{2} \delta(x-(y+|y|+1)) + \tfrac{1-\eta}{2} \delta(x-(y-|y|-1)) +\eta\delta(x-y),
    \end{split}
\end{equation}
for some small $\eta>0$. In this case $y$ is always the median of $g(x|y)$, and $D(g)|_y = (1-\eta)(|y|+1)\geq 1-\varepsilon$. However, we have:
\begin{equation}
    \begin{split}
        h(y) &= \begin{cases} \eta\,\,\mathrm{for}\,\, -\tfrac{1}{2}<y<\tfrac{1}{2}\\
        0\,\,\mathrm{elsewhere} \end{cases},
    \end{split}
\end{equation}
which is not a normalised probability density function since $\int h(y) dy = \eta$, and moreover has mean absolute deviation $D(h) = \tfrac{\eta}{4}$, vanishing in the limit of small $\eta$. While this example is clearly un-physical, it illustrates that nontrivial assumptions about the form of $g(\varepsilon|\mu)$ are necessary to lower-bound $D(p)$ in terms of $D(g)$.



%